\documentclass[10pt]{article}
\usepackage{amsmath, amssymb, graphicx, color, setspace, authblk}
\usepackage[margin=1in]{geometry}

\newcommand{\nn}{\nonumber}

\renewcommand{\eqref}[1]{Eq.~(\ref{#1})}
\newcommand{\eqs}[2]{Eqs.~(\ref{#1}) and (\ref{#2})}
\newcommand{\ev}[1]{E\left[ #1 \right]}
\newcommand{\evb}[1]{\left\langle #1 \right\rangle}
\newcommand{\evo}[2]{E_{#2}\left[ #1 \right]}
\newcommand{\fig}[1]{Fig.~\ref{fig:#1}}
\newcommand{\figs}[2]{Figs.~\ref{fig:#1} and \ref{fig:#2}}
\newcommand{\bigo}[1]{\mathcal{O}\left(#1\right)}

\newcommand{\given}{\ | \ }

\newcommand{\ox}{f}
\newcommand{\spd}{c}
\newcommand{\wid}{l}
\newcommand{\pop}{\rho}
\newcommand{\size}{L}
\newcommand{\mig}{D}
\newcommand{\bt}{t}%backwards time
\newcommand{\xc}{x_c}
\newcommand{\tcon}{t_\text{con}}
\newcommand{\mige}{\mig_\text{eff}}
\newcommand{\val}{\phi}
\newcommand{\ld}{\Delta}
\newcommand{\pneut}{p_\text{neut}}
\newcommand{\ibd}{\psi}
\newcommand{\tcoal}{T}
\newcommand{\mut}{\mu}

\title{Ancestry in adapting, spatially-extended populations}
\author[1]{Daniel B.~Weissman\footnote{daniel.weissman@emory.edu}}
\affil[1]{Dept.~of Physics, Emory University, Atlanta, GA 30322}
\date{}

\begin{document}

\maketitle

\section*{Abstract}

Selective sweeps affect neutral genetic diversity through hitchhiking. 
While this effect is limited to the local genomic region of the sweep in panmictic populations,
we find that in spatially-extended populations the combined effects of many unlinked sweeps
can affect patterns of ancestry (and therefore neutral genetic diversity) across the whole genome.
Even low rates of sweeps can be enough to skew the spatial locations of ancestors such that 
neutral mutations that occur in an individual living outside a small region in the center of the range
have virtually no chance of fixing in the population.

\section*{Introduction}

In large populations even a fairly low rate of selective sweeps is sufficient to reduce diversity across most of the genome
via hitchhiking \cite{gillespie2000, weissman2012}.
Most modeling of the effects of hitchhiking on diversity has considered well-mixed populations.
However, the effects are potentially quite different in spatially-extended populations,
because instead of quickly fixing through logistic growth, sweeps must spread out in a spatial wave of advance
over the whole range \cite{fisher1937}.
\cite{barton2013} recently showed that this increase in the time to sweep tends to reduce
the size of the genomic region over which diversity is depressed by a sweep.

While the effect of sweeps on genetic diversity at linked loci is therefore reduced by spatial structure,
we show here that collective effect of sweeps on the diversity at \textit{unlinked} loci can be much
stronger than in panmictic populations.
Surprisingly, this effect is dependent on the geometry of the range -- it only appears for realistic range shapes with relatively well-defined central regions,
not for the perfectly symmetric idealizations of ring-shaped and toroidal ranges often used in theoretical models.
In particular, we find that probability of fixation of an allele can be strongly position-dependent, 
with alleles near the center of the range orders of magnitude more likely to fix than those at typical locations.
This can produce a false signal of a population expansion (in number, not space) even at loci on chromosomes where no adaptation is taking place.
The basic mechanism driving this effect is that sweeps tend to arrive at each location from the direction of the center of the range,
and so bias the ancestry back towards the center.

\section*{Methods}

We wish to find the expected number of 
copies that an allele found in an individual at spatial position $\mathbf x$ will leave far in the future, i.e., its reproductive value \cite{barton2011}, which we denote $\val(\mathbf x)$.
Equivalently, $\val(\mathbf x)\pop(\mathbf x)$, where $\pop$ is the population density,  is the probability density of
a present-day individual's  ancestor being at location $ \mathbf x$ at some time in the distant past. 
\cite{maruyama1970} showed that in the absence of selection, 
$\val(\mathbf x) \equiv 1$ regardless of the details of the population structure, as long as dispersal does not change expected
allele frequencies. 
Here we show that this result does not extend to populations undergoing selection.
Populations living in perfectly symmetric ranges (circles in one dimension, tori in two)  necessarily have $\val(\mathbf x) \equiv 1$, but when this symmetry is broken,
recurrent sweeps can create a substantial imbalance, making $\val$ much higher in a small region in the center of the range and decreasing it everywhere else.

\section*{Model}
We consider a population with uniform, constant density $\pop$ distributed over a $d$-dimensional range with radius $\size$, with uniform local dispersal with diffusion constant $\mig$. 
We assume that selective sweeps with advantage $s$ occur in the population at a rate $\Lambda$ per generation, originating at points uniformly distributed over time and space.
As long as the density is sufficiently high ($\pop \gg \left(s / \mig\right)^{d/2} / s$, \cite{nagylaki1978, barton2013}), they will spread roughly
deterministically in waves with speed $\spd \approx 2\sqrt{\mig s}$ with characteristic wavefront width $\wid \approx \sqrt{\mig/s}$ \cite{fisher1937},
which we take to be much smaller than the range size, $\wid \ll \size$. 
We assume that $\Lambda$ is low enough compared to the frequency of outcrossing, $\ox$, and the average number of crossovers per outcrossing, $K$, that the waves do not interfere with each other. 
The definitions of symbols are collected in Table \ref{tab:defs}.

\subsection*{One and two dimensions}
We consider both one-dimensional ranges (lines with length $2\size$) and two-dimensional ranges. 
In two dimensions, the shape of the range will have some effect on many of our results; 
however, as long as the shape is fairly ``nice'', with a clear center and single characteristic length scale $\size$, this effect will be 
modest.
We will therefore ignore it for simplicity. 
For our purposes, the main difference between one and two dimensions will be in the density of individuals a distance $x$ from the center, $\pop(x)$. 
Since we are assuming a uniform spatial density, in one dimension this is just $\pop$, a constant. In two dimensions, however, we must account for the fact that
there is more area at larger $x$, and thus $\pop(x)\approx2\pi x \pop$. (Obviously, $\pop(x>\size)=0$ in both one and two dimensions.)

\section*{Results}

A sweep at a tightly-linked locus a genetic map distance $r \ll s$ away pulls a lineage a distance that is approximately exponentially distributed with mean $\spd/r$, going backwards in time. 
This is only approximate, because actually there is an upper cutoff at the distance to the origin of the sweep \cite{barton2013}. 
For sweeps at loosely-linked loci with $r\gg s$, the lineage is only pulled an expected distance $\spd/2r$ (see Appendix A). 
However, since the average displacement still only falls off like $1/r$, and since there is an upper cutoff on the effect of sweeps as $r\to0$, 
the total average displacement for a typical locus can be dominated by the effect of the many unlinked sweeps rather than the few linked ones, assuming that sweeps are 
uniformly distributed over the genome, and that the genome is long. We will make the approximation that the expected displacement is solely due to unlinked sweeps, with $r=\ox/2$;
we will consider the additional effects of the rare, tightly-linked sweeps below.

 For a lineage a distance $x$ from the center, there is an excess of approximately 
 $\sim \Lambda x/\size$ sweeps per generation pulling it back toward the center, each of which pulls it an expected distance $\spd/\ox$. 
 (Note that the effect of the upper cutoff on the displacement from these sweeps is negligible as long as $\size \gg \spd/\ox$.)
 The expected distance from 
the center therefore decays exponentially (backwards in time) like 
\begin{equation}
\overline x \approx x_0 \exp\left(-\frac{\Lambda\spd}{\size\ox}\bt\right).\label{Ex_t}
\end{equation}
This implies that there is a time $\tcon$ before which individuals' ancestors are unlikely to be found outside the center of the range, with 
\begin{equation}
\tcon = \frac{\size\ox}{\Lambda\spd}.
\end{equation}

This deterministic move back to the center is opposed by dispersal, and also by the effect of occasional tightly-linked sweeps which pull the lineage a 
distance $\sim \size$, effectively randomizing its position. The balances between these forces means that the ancestry of the population is not completely 
concentrated at the center of the range, but is instead distributed around it in some region of size $\sim \xc$. 

\subsection*{Balance with dispersal}

If tightly-linked sweeps are relatively rare, either because the overall rate of sweeps is low or because the focal locus lies in a region of the genome that is not 
undergoing much adaptation, the main balance will be between the diffusive effect of dispersal and the pull of unlinked sweeps. 
In this case, the position of the ancestry is an Ornstein-Uhlenbeck process.
The stationary distribution is therefore normal and concentrated in the center of the range according to:
\begin{equation}
\val(x) \propto \exp\left(-\frac{x^2}{2\xc^2}\right),\text{ with }\frac{\xc}{\size}=\sqrt{\frac{\ox\wid}{2\Lambda\size}}.\label{migval}
\end{equation}
If $\xc \ll \size$, then the reproductive value of an individual at the center of the range can be orders of magnitude higher than than one at a typical distance $\sim L/2$ from the center (\fig{ancdist}).

From \eqref{migval}, we see that the ancestral range will be substantially reduced by selection if the rate of sweeps per sexual generation is greater than the ratio of the cline width
to the range size: $\Lambda/\ox > \wid/\size$.
It is unclear what ranges these ratios take in natural populations. 
$\Lambda/ (\ox K)$ is unlikely to be much more than $\bigo{1}$ \cite{weissman2012}, but in organisms with many chromosomes (large $K$), $\Lambda/\ox$ may be substantial.
Looking at the right-hand side of the inequality, modeling sweeping alleles by waves spreading across the range necessarily requires $\wid / \size \ll 1$, so even small values of 
$\Lambda/\ox$ may be enough to distort the distribution of ancestry.
Surprisingly little is known about typical values of $\wid$ for the waves of advance of sweeping alleles in nature, 
but it seems plausible that for many species it should be much smaller than the total species range \cite{fisher1937}.
For the spread of insecticide resistance in \textit{Culex pipiens} in southern France, the width of the wave of advance was $\sim 20 \text{ km}$ \cite{lenormand1999}, 
much smaller than the global scale of the species range, but the dynamics were more complex than a simple selective sweep \cite{labbe2007}.
Much more is known about the width of stable clines and hybrid zones, which are frequently much smaller than species ranges \cite{barton1985}.
To the extent that the selection maintaining them is comparable in strength to the selection driving sweeps,
these should have roughly the same width as the wavefronts.

\subsection*{Balance with tightly-linked sweeps}

Finding the balance between concentrating effect of unlinked sweeps and the randomizing effect of tightly-linked sweeps is slightly trickier. 
Indeed, finding an exact expression for $\val(x)$ is intractable.
However, we can find an approximate expression by using the fact that the mean squared displacement of the ancestral lineage due to linked 
sweeps is dominated by rare very tightly-linked sweeps rather than the many loosely-linked ones \cite{barton2013}.
This suggests that for large $x$, the probability that an individual's ancestor was farther than $x$ from the center at time $t_0$ in the distant past is 
roughly just the probability that a single very tightly-linked sweep pulled it there at some time within $\sim \tcon$ generations of $t_0$.
Since the distance that a sweep at recombination fraction $r$ pulls the lineage goes like $1/r$, the rate of sweeps close enough on the genome to 
pull the ancestry a distance of at least $x$ falls off like $1/x$. 
Therefore, the probability of finding the ancestry at a distance of at least $x$ should also fall off like $1/x$; 
the probability density of being exactly at $x$, $\val(x)\pop(x)$, should then fall off like $1/x^2$.

In the appendix, we calculate this more formally, and find
\begin{equation}
\val(x)\pop(x) \approx \frac{2\size\left(1-(x/\size)^d\right) }{K x^2}\text{ for }x\gg \xc=2\size/K.\label{linked_tail}
\end{equation}
The factor $1-(x/\size)^d$ (where $d=1\text{ or }2$ is the dimension of the habitat) reflects the fact that for very large $x$, $x\sim L$, most sweeps start at distances less than $x$ and cannot pull the lineage that far from the center.
For $x\ll\xc=2\size/K$, lineages will tend to experience many sweeps pulling them distances greater than $x$ in time $\sim \tcon$,
so the approximation used to derive \eqref{linked_tail} breaks down;
for these small values of $x$, the randomizing effects of moderately-linked sweeps smooth out $\val(x)$ and make it roughly constant.

Barton et al.~describe the randomizing effect of tightly-linked sweeps by ``$\mige$,'' an effective dispersal rate, with $\mige\approx \frac{16\size\Lambda}{3\wid K
\ox }\mig$ (Eq.~(9) of \cite{barton2013}). Comparing \eqs{migval}{linked_tail}, however, we see that their effect cannot simply be described as an increase in the dispersal rate, 
since they create a much longer tail in the spatial distribution of ancestry. 
Because of this, it is possible that while the bulk of the distribution of ancestry is determined by a balance between unlinked sweeps and dispersal, with linked sweeps too rare to make a difference,
linked sweeps make the dominant contribution to the tails of the ancestry distribution (\fig{logancdist}).

\subsection*{Combining dispersal and tightly-linked sweeps}

Combining \eqs{migval}{linked_tail}, we see that unlinked sweeps reduce the effective size of the ancestral range by a factor $\xc/\size$:
\begin{equation}
\frac{\xc}{\size} \approx \max\left\{\sqrt{\frac{\ox\wid}{2\Lambda\size}}, \frac{2}{K}\right\}\label{center_scaled}.
\end{equation}
For typical numbers of chromosomes $K$, it would seem that ancestry could be concentrated by about an order of magnitude. 
However, the result $2/K$ was derived 
under the assumption that sweeps are distributed uniformly across the genome. 
If, on the other hand, adaptation is mostly occurring in just a few genes, the rest of the genome
will not experience any tightly-linked sweeps, and ordinary dispersal will be the only force counteracting the concentration,
meaning that the effect could potentially be much stronger. This has the surprising implication that 
selection can have a stronger effect on some features of the spatial distribution of ancestry at far-away loci than at those nearby.

\subsection*{Effect on diversity}

While the effect of recurrent sweeps on neutral diversity can be quite large, detecting the effect in data from real populations may be tricky. 
It might seem to be indistinguishable from a range expansion in the absence of time-series data, but there is a simple way to tell them apart: 
under recurrent sweeps, there is no serial founder effect reducing diversity away from the center.
One way to see this is by looking at isolation by distance. 
The probability $\ibd(x)$ that two individuals separated by a distance $x$ are genetically identical can be written in terms of the neutral mutation rate $\mut$ and  their coalescence time $\tcoal$ as
\begin{equation}
\ibd(x) = \ev{e^{-2\mut \tcoal} \given x}.\label{ibdT}
\end{equation}

For $x$ large compared to the size of a single deme (i.e., the spatial scale over which individuals interact within a generation) and loci far on the genome from any recent sweeps, 
there are two simple regimes for \eqref{ibdT}.
If $x\ll\xc$, then we expect that the pull due to sweeps should be unimportant, and $\ibd(x)$ is just given by the neutral value,
$\ibd(x) \propto x^{(1-d)/2}e^{-\sqrt{\mut/\mig}x}$ \cite{barton2002a}, which says that
the probability of identity falls off rapidly with distance.
On the other hand, larger values of $x$ are quickly collapsed by the pull of sweeps in time $\sim \log(x)/\tcon$, so we expect that 
$\ibd$ should be of the form $\ibd(x)\propto x^{-2\mu\tcon}$.
A detailed calculation in Appendix C confirms that this is true for $x \gg \xc \sqrt{2+4\mut \tcon}$.
The probability of identity thus has a long tail in distance -- 
individuals at opposite sides of the range (separated by $\approx2L$) 
are nearly as related as individuals separated by, say, $L/2$.
Notice that $\ibd$ does not depend on from where in the range we sampled the pair of individuals.
This implies that, while reproductive value is concentrated in the center of the range, 
genetic diversity is more evenly spread, distinguishing this scenario from a range expansion.

Above, we have ignored loci that are close to recent sweeps. 
If we are considering large enough loci so that $\mut  \tcon \gg 1$, then usually only these 
recently swept regions will be identical between individuals from different parts of the range.
In this case, because each sweep causes coalescence between individuals separated by a large distance $x$
over a region of genome with length $r \propto 1/x$ \cite{barton2013}, $\ibd$ should still have a long tail,
but with an exponent that is independent of the population parameters, $\ibd\propto 1/x$ (see Appendix C.1).
This characteristic exponent is another effect of rare, tightly-linked sweeps that cannot
be accounted for by any effective dispersal rate $\mige$.

\section*{Discussion}

Because selection and demography are often difficult or impossible to measure directly in natural populations, both are typically inferred from patterns of genetic diversity.
This inference can be  difficult, because the two processes can produce similar signals. 
For instance, both purifying selection and population expansion tend to produce site frequency spectra with a relative excess of rare alleles.
In order to tease apart the two factors, demography is often first inferred using data from loci that are thought to be neutral, 
and then the answer is used to infer the pattern of selection at  the remaining loci.
However, in order for the demography to be inferred correctly, this method requires that the first set of loci be not just neutral, 
but also unaffected by selection at linked loci. 
Typically, this is done by using loci that are far from sites where selection is thought to have been important (e.g., \cite{sattath2011}).
Our results suggest that this may be problematic in spatially-structured populations -- 
even diversity at these loci may be strongly affected by unlinked sweeps.

\subsection*{Geometry, not topology, of range is important}

\begin{figure}
\begin{center}
 \includegraphics[width=6in]{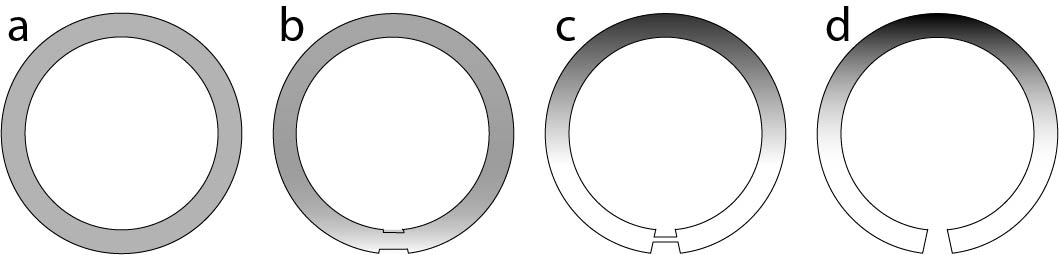}
 \caption{\label{fig:circles} The distribution of ancestry depends smoothly on the shape of the range as it deforms from a perfectly symmetric
 circle (a) to a curve with endpoints (d). 
 Shading represents the reproductive value of each location, from high (dark) to low (light). 
In (a), the ancestry is necessarily evenly distributed.
Slight asymmetries in the shape of the range introduce slight differences in the distribution of ancestry (b).
When the range has a well-defined middle (d), the ancestry is concentrated there;
including a weak connection between the end of the range (c) does not change this much.
}
\end{center}
\end{figure}

Our results might seem to show that the genetic diversity in a population depends sensitively on the topology of the range and can therefore change drastically
as the result of small perturbations to the environment. 
For example, a circular range (which has no concentration of ancestry since it is perfectly symmetric) can be transformed into a linear one (with very 
concentrated ancestry) by removing a single point.
However, this is a misleading interpretation. 
In fact, a ``circular'' range is an annulus with radius large compared to its thickness (\fig{circles}a).
A small perturbation that slightly reduces the population in one part of the range will only have a correspondingly small effect on the distribution 
of ancestry (\fig{circles}b), and the bias of the ancestry ancestry increases smoothly as the perturbation grows (\fig{circles}c), 
until the annulus is completely pinched off (\fig{circles}d).
More generally, the common-sense intuition that the pattern of diversity should not depend on the details of the shape of the range is correct.
All that matters is the extent to which there is a ``central region''
that sweeps tend to pass through on their way to dominating the population.

\subsection*{Extensions} 

We have focused on a very simple population model. 
Here we consider several possible modifications.
First, we have assumed that the density $\pop$ is constant in time.
If density fluctuations typically occur on timescales longer than $\tcon$,
this approximation should be accurate, and if they are rapid compared to the sweep time $\size/\spd$
they should average out, but it is unclear how fluctuations on moderate timescales
should interact with dynamics discussed here.

We have also neglected the possibility of rare long-range dispersal.
Tightly-linked sweeps already effectively produce occasional long-range jumps in the  ancestry of neutral sites,
so adding long-range dispersal might not have a large direct effect,
but it is likely to have dramatic effects on how sweeps spread \cite{hallatschek2014},
and therefore a large indirect effect on the hitchhiking dynamics.
It is not at all clear what this effect should be -- on the one hand, 
the sweeps will spread faster, increasing their pull,
but on the other hand, the direction of that pull may be less reliably towards the center.

We have also neglected the possibility that many sweeps may be ``soft'',
starting from multiple alleles \cite{hermisson2005}.
If these alleles typically descend from a recent single ancestor, i.e,
are concentrated in a small region at the time when they begin to sweep,
then the results should be essentially unchanged, with the possible exception of
the coalescent effects of tightly-linked sweeps (Appendix C.1).
The same should be true if sweeps are ``firm'', i.e., multiple mutant lineages contribute to each sweep,
but the most successful one typically colonizes most of the population.
But sweeps in which many widely-spread mutations contribute equally would likely 
not consistently concentrate ancestry in space.

We have focused on the effect of sweeps on neutral variation,
 but they will of course also affect selected alleles.
 Most obviously, if recombination is limited they will  interfere with each other \cite{martens2011}. 
 They will interfere even more strongly with weakly-selected variants.
We are currently preparing a manuscript addressing these issues.
It is also important to consider how the concentration of reproductive value
interacts with spatially-varying selection. It seems plausible that it would reduce 
the potential for local adaptation and thereby limit population ranges.

\section*{Acknowledgements}

This work would not have been possible without Nick Barton's generous assistance at all stages.

\clearpage

\section*{Tables}
\begin{table}[!ht]
\caption{
\bf{Symbol definitions}}
\begin{tabular}{|c|l|}
\hline
\textbf{Symbol} & \textbf{Definition}\\
\hline
$\pop$ & Density of individuals\\
$\size$ & Radius of range\\
$\mig$ & Dispersal constant\\
$\ox$ & Frequency of outcrossing\\
$K$ & Expected number of crossovers per outcrossing\\
$s$ & Selective advantage of adaptive alleles\\
$\Lambda$ & Rate of adaptive substitutions\\
$\spd\approx 2\sqrt{\mig s}$ & Expected rate of advance of a sweeping beneficial allele\\
$\wid\approx \sqrt{\mig/s}$ & Characteristic width of the wave of advance of a sweeping beneficial allele\\
\hline
\end{tabular}
\begin{flushleft} The definitions of the main symbols used in the text.
\end{flushleft}
\label{tab:defs}
 \end{table}

\clearpage

 \section*{Figures}
 
 \begin{figure}[bh]
 \includegraphics[width=6in]{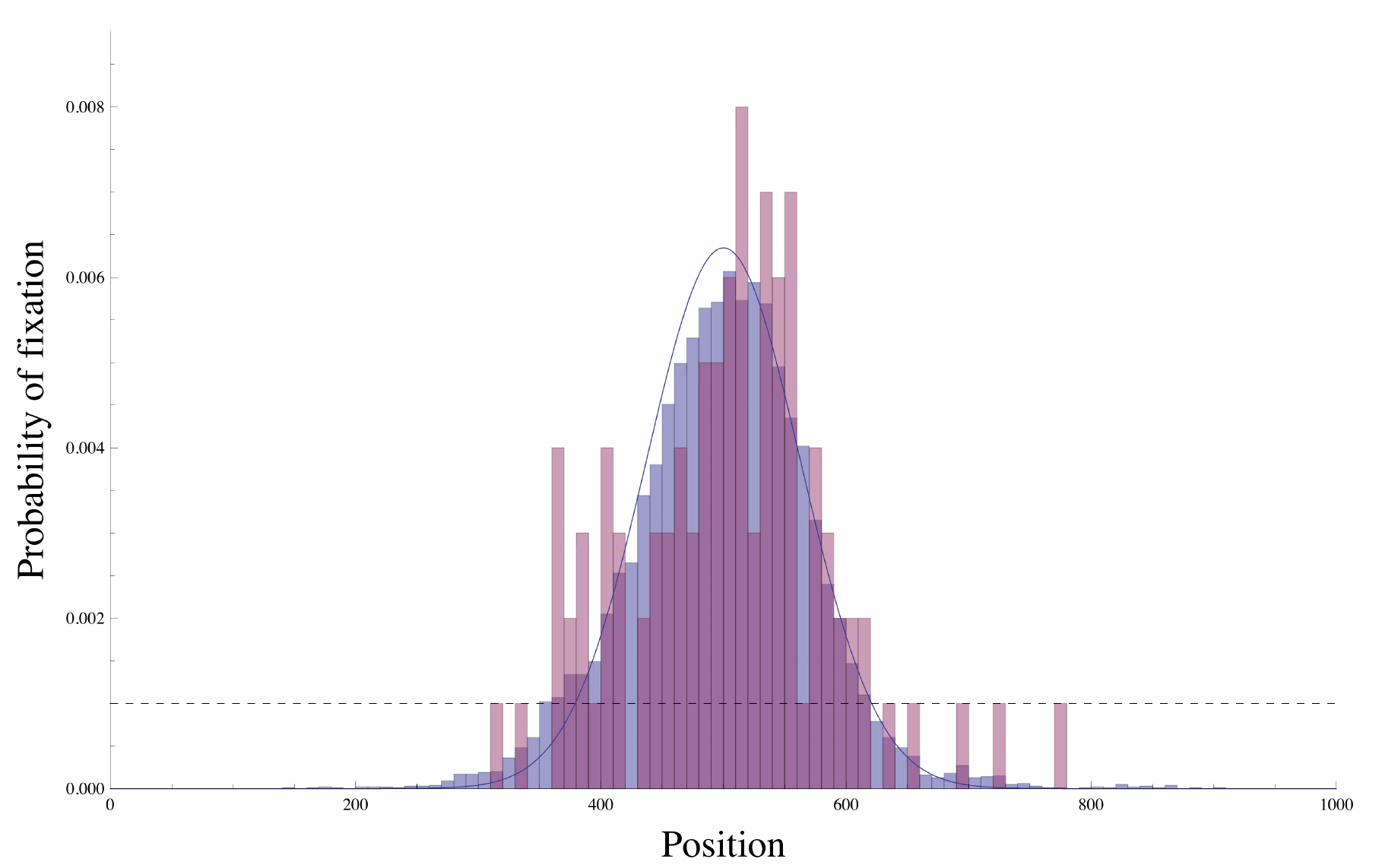}
 \caption{\label{fig:ancdist} Unlinked sweeps concentrate ancestry in the center of the range. The plot shows the long-term probability of finding the ancestor of a neutral allele at a given deme in a linear range. The purple histogram shows full-population forward simulations, the blue histogram shows approximate backwards simulations, and the blue curve shows the predicted distribution, \eqref{migval}. The dotted black line shows a uniform distribution. Parameters are $\size=500, \pop=100, s=0.05, \mig=0.125, \Lambda=0.1, \ox=1, K=100$; see Appendix D for simulation methods.}
 \end{figure}
 
\clearpage

\begin{figure}
\begin{center}
 \includegraphics[width=\textwidth]{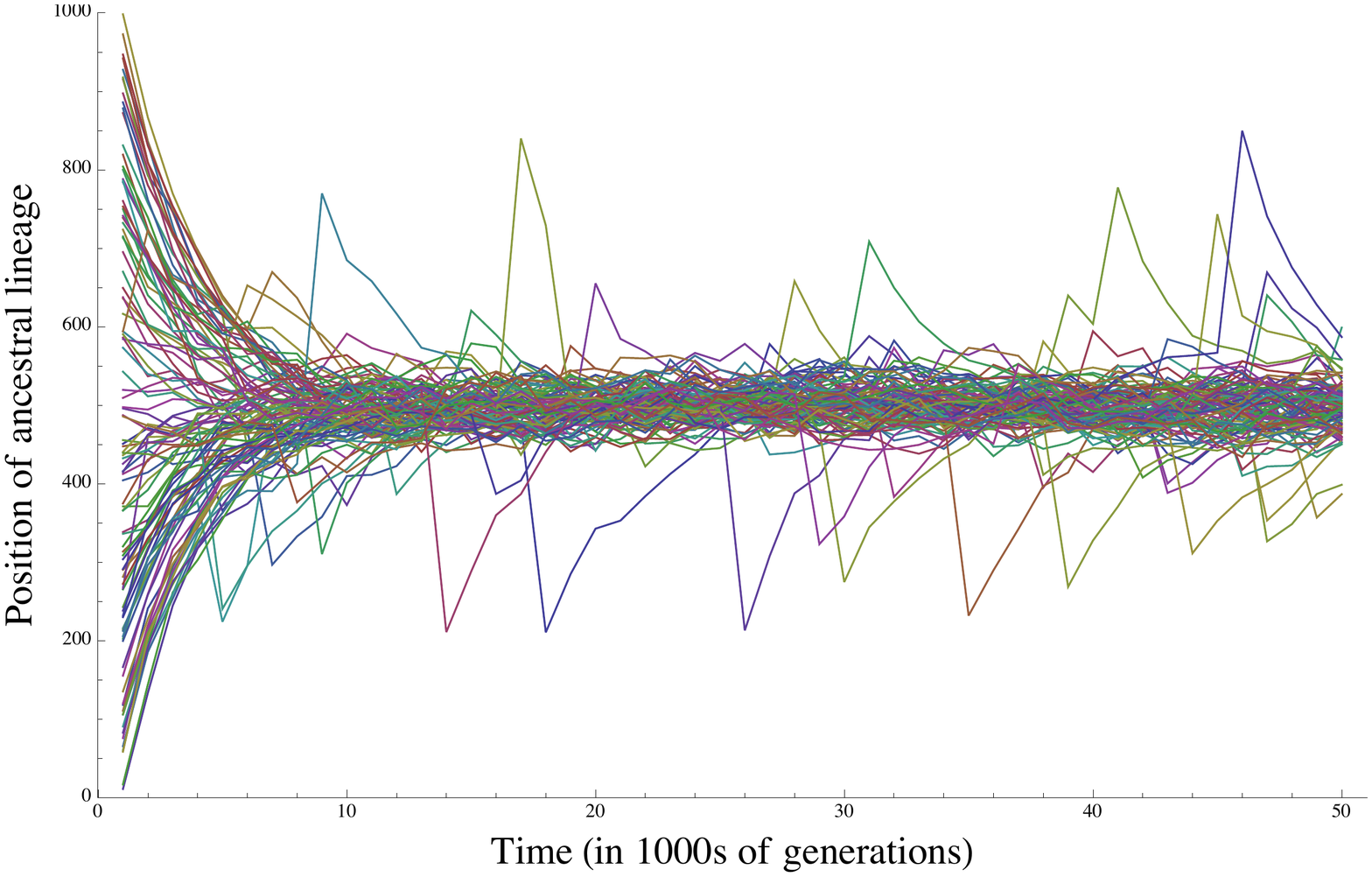}
 \caption{\label{fig:trace} Tracing the ancestry of an individual back through time. 
 Individuals throughout the range rapidly trace their ancestry at a neutral locus back to a small region in the center of the range.
 The width of this region is determined by strength of dispersal.
 The occasional excursions out of the center are due to hitchhiking on tightly-linked sweeps. 
  Parameters are $\size=500, s=0.05, \mig=0.125, \Lambda=1, \ox=1, K=300$.
  Each curve is an independent simulation.}
 \end{center}
\end{figure}

\clearpage

\begin{figure}
\begin{center}
 \includegraphics[width=6in]{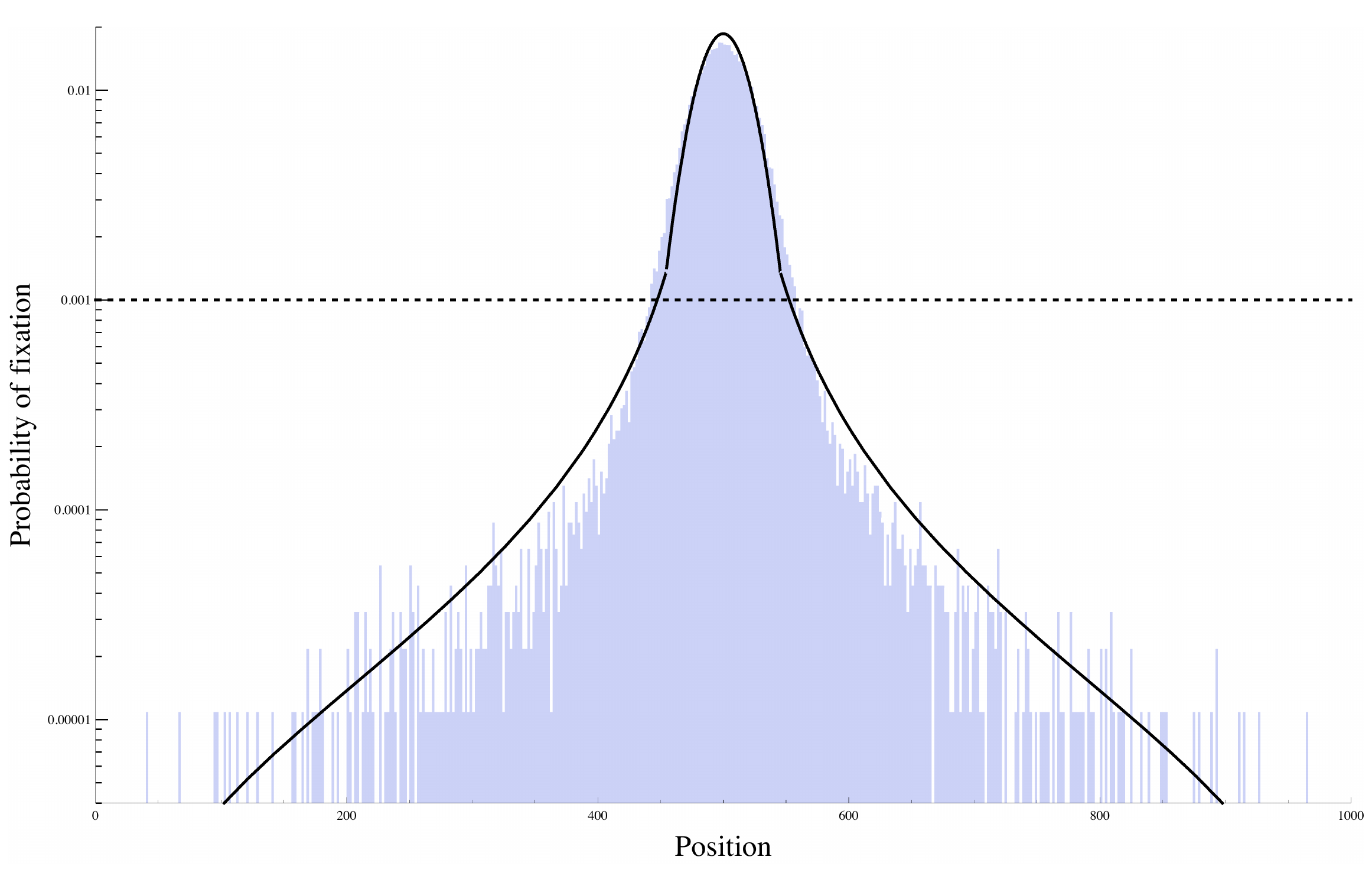}
 \caption{\label{fig:logancdist} The long-term probability of finding the ancestor of a neutral allele at a given deme in a linear range. 
 The blue histogram shows approximate backwards simulations, the dotted black line shows a uniform distribution, and the blue curve shows the predicted distribution, \eqs{migval}{linked_tail}. 
 Typical loci are in the center of the range, in a balance between the pull of unlinked sweeps and dispersal (\eqref{migval}), but the (small) probability that the ancestor is found 
 outside the center of the range is dominated by the probability that there has recently been a tightly-linked sweep (\eqref{linked_tail}). 
 Parameters are as in \fig{trace}.}
 \end{center}
\end{figure}

\clearpage

\clearpage

 \bibliography{references}
 
 \appendix
 
 \section{Calculating the ``pull'' of an loosely-linked sweep}
 
 We would like to find the expected spatial displacement of a lineage caused by an loosely-linked sweep, tracing backwards in time. 
 To do so, suppose that we sample an allele in a present-day individual in the middle of a very large one-dimensional range, and that a long time ago a selective 
 sweep occurred at a locus a recombination fraction $r$ away from the focal allele, starting a very long distance away from our sample. 
 We wish to find the expected location of the ancestor of the sampled allele before the sweep began. 
 Let $p(x,\tau)$ be the probability density for finding the ancestor at location $x$ $\tau$ generations in the past, with $x=0$ corresponding to the present location.
 We want to find 
 \begin{align}
 \lim_{\tau\to\infty} \evb{X}  \equiv & \lim_{\tau\to\infty}\int dx \, x p(x,\tau)\nn\\
 = & \int d\tau \int dx \, x \partial_\tau p(x,\tau).\label{dxdef}
 \end{align}

 To find $\partial_\tau p$, first define $p_i(x,\tau)$ as the probability density that the ancestor was at location $x$ and in genetic background $i$, where $i=0$ is the 
 ancestral genetic background, and $i=1$ is the background with the allele that swept. (Note $p=p_0+p_1$.) 
 If we define $u(x,\tau) \equiv u_1(x,\tau)$ and $u_0(x,\tau)$ to be the frequencies of the sweeping allele and the background allele, respectively, with $u_1+u_0=1$, 
 $p_i$ satisfies the partial differential equation
 \begin{equation}
 \partial_\tau p_i = r(u_i p_{1-i}-u_{1-i}p_i)+\mig\partial_x (\partial_x p_i -2p_i\partial_x \log u_i). \label{dpidT}
 \end{equation}
Technically, in models with discrete generations, \eqref{dpidT} only applies when the recombination rate per generation is small, but we will use it for unlinked loci anyway.
 
The equivalent of linkage disequilibrium in this system is $\ld \equiv u_0 p_1-u_1 p_0$; we expect it to be small for large $r$. Using $\ld$ to change variables back to $p$, \eqref{dpidT} becomes
\begin{align}
\partial_\tau p  = & \mig \partial_x\left(\partial_x p - 2 \frac{\partial_x u}{u(1-u)}\ld\right)\label{dlocdT}\\
\partial_\tau \ld  = & -r\ld+\mig\partial_x^2 \ld -(\partial_\tau u +\mig\partial_x^2 u)p+2(2u-1)\mig\partial_x\left(\frac{\partial_x u}{u(1-u)}\ld\right)+2\mig\frac{\partial_x u}{u(1-u)}\ld\label{dLDdT}
\end{align}
 
 Plugging \eqref{dlocdT} into \eqref{dxdef}, we have
  \begin{align}
 \lim_{\tau\to\infty} \evb{X} & = \mig \int d\tau \int dx \, x \partial_x\left(\partial_x p - 2 \frac{\partial_x u}{u(1-u)}\ld\right)\nn\\
 & = 2\mig \int d\tau \int dx \frac{\partial_x u}{u(1-u)}\ld,\label{dxLD}
 \end{align}
 where we have used integration by parts and the fact that $p(\pm \infty,\tau) = 0$. It now remains to find an expression for $\ld$. 
 \eqref{dLDdT} is quite complicated, but for large $r$ we will have $\ld \ll p$ and the dominant balance will be between the first and third terms on the
 right-hand side, giving
 \begin{equation}
 \ld \approx -\frac{1}{r}(\partial_\tau u+\mig \partial_x^2 u)\pneut,\label{LDapprox}
 \end{equation}
 where $\pneut$ is the value of $p$ ignoring the perturbation caused by the sweep, i.e., $\pneut = \frac{1}{\sqrt{4\pi\mig\tau}}\exp\left(-\frac{x^2}{4\mig\tau}\right)$.
 We can simplify this further by noting that $u$ solves the differential equation $\partial_\tau u +\mig \partial_x^2 u = -s u(1-u)$. (Recall that $\tau$ is
 backwards time.) Using this relation and substituting \eqref{LDapprox} into \eqref{dxLD}, we have
 \begin{equation}
  \lim_{\tau\to\infty} \evb{X} = \frac{2\mig s}{r} \int d\tau \int dx \, \partial_x u \, \pneut.\label{dxpneut}
  \end{equation}
 
 Recall that we are interested in the effect of a long-past sweep. Let $\tau_0$ be the time at which the wave of advance passed the point where we sampled
 the allele; we will take $\tau_0$ to be extremely large. 
 At time $\tau_0$, $\pneut$ has width $\sim \sqrt{\mig \tau_0}$, so the wave crosses the region where the ancestor might have lived in a time 
 $\sim \sqrt{\mig \tau_0}/\spd \ll \tau_0$, and the 
 integral in \eqref{dxpneut} is dominated by times $\tau$ in the approximate range $|\tau-\tau_0| \lesssim \sqrt{\mig \tau_0}/\spd$. 
 Since $\tau$ does not vary by much (proportionately) in this interval, $\pneut(x,\tau)\approx \pneut(x,\tau_0)$ is approximately constant in $\tau$.
 Using this approximation in \eqref{dxpneut} yields
 \begin{align}
   \lim_{\tau\to\infty} \evb{X} \approx & \frac{2\mig s}{r} \int dx \, \pneut(x,\tau_0)   \int d\tau \, \partial_x u \nn\\
    = & \frac{2\mig s}{r} (1) \left(-\frac{1}{\spd}\right)\nn\\
    = & -\frac{c}{2r}.\label{evdx}
 \end{align}
  Note that this result did not depend on the form of $\pneut$, only that it was approximately constant in time; in particular, it also holds if the ancestry settles down to a 
  stationary distribution, as in \eqref{migval}.
 
 \subsection{Other kinds of loosely-linked sweep}
 
Above, we have assumed that the sweeping allele spread according to the FKPP equation, $\partial_\tau u +\mig \partial_x^2 u = -s u(1-u)$,
which describes an allele with a constant selective advantage $s$.
However, the allele may have a varying selective advantage if, for instance, dominance or frequency-dependent effects are important,
or if there is environmental variation. 
More generally, the changing allele frequency is described by 
\begin{equation}
\partial_\tau u +\mig \partial_x^2 u = -s f(u,x,\tau)u(1-u)\label{genwave}
\end{equation}
for some function $f$.

Otherwise, the derivation of the expected displacement is the same as above, and we have
\begin{equation}
   \lim_{\tau\to\infty} \evb{X} \approx  \frac{2\mig s}{r} \int dx \, \pneut(x,\tau_0)   \int d\tau \, f(u,x,\tau) \partial_x u.
\end{equation}
 Assuming that $f$ is such that $u(x,\tau)$ is still a traveling wave moving at some speed $\spd$, we can change variables in the second integral to obtain:
 \begin{equation}
   \lim_{\tau\to\infty} \evb{X} \approx  -\frac{2\mig s}{r\spd} \int dx \, \pneut(x,\tau_0)   \int_0^1 du\ \, f(u,x,\tau(x,u)).
\end{equation}

\section{Effect of tightly-linked sweeps}
 
 We wish to calculate $\val(x)$ for large $x$, including the effect of occasional tightly-linked sweeps. 
 It is easiest to consider $\int_x^L dy \, \pop(y)\val(y)$, which we can think of as the probability that at some time $t_0$ in the distant past, 
 the ancestor of a present-day individual was at a distance greater than $x$ from the center. 
 For large $x$, we expect that this is dominated by the probability that it was pulled there by a 'recent' tightly-linked sweep  $t$ generations
 `before' $t_0$ (i.e., $t$ generations closer to the present), with $t$ not too large. 
This sweep must have pulled the lineage out to a distance of at least $xe^{t/\tcon}$ for it still to be at a distance of at least $x$ $t$ generations
`later', and therefore the sweep must have originated a distance $z>xe^{t/\tcon}$ from the  center. 
Given that it did, the probability that it pulled the lineage out far enough is $\exp\left[-\frac{r}{c}xe^{t/\tcon}\right]$.
Putting this all together, and using that the density of sweeps per generation per unit map length per distance (or area in two dimensions) at distance $z$ from the 
center and genetic map distance $r$ from the focal locus is $2\Lambda/(\ox K \size)$ (or $4\Lambda z/(\ox K\size^2)$ in two dimensions),
 the expected number of sweeps that would have left the lineage more than $x$ from the center at time $t_0$ is
\begin{align}
\int_x^L dy \, \pop(y)\val(y) & \approx \frac{2\Lambda}{\ox K\size^d}\int_0^{\tcon \log(\size/x)}dt \int_{xe^{t/\tcon}}^\size dz \, (2z)^{d-1} \int dr\, \exp\left[-\frac{rx}{c}e^{t/\tcon}\right]\nn\\
& = \frac{2\size}{K x}\times\begin{cases}1-(1+\log (\size/x))x/\size & \text{for } d=1\\ (\size-x)^2/\size^2 & \text{for } d=2.\end{cases}\label{linkCDF}
\end{align}
Taking the derivative of both sides of \eqref{linkCDF} with respect to $x$ gives the probability density, \eqref{linked_tail}.

Note that \eqref{linkCDF} approximates the probability that there is at least one tightly-linked sweep  by the expected number of such sweeps,
so it is only valid when the right-hand side is small, $x\gg 2\size/K$.
It also obviously typically breaks down as $x$ approaches $\size$ and the particular geometry of the habitat begins to matter.

 \section{Isolation by distance}
 
 We wish to find the probability $\ibd(x)$ that a pair of lineages a distance $x$ apart will be identical at a neutral locus.
Let us assume that the locus is far from any recent sweeps. (We relax this assumption below.)
Then tracing the ancestry back in time, the separation $X_\tau$ between them can be approximated by a Brownian motion, 
with diffusion constant $2\mig$ (since it combines the motion of both lineages), and with the lineages moving together
at a mean rate of $\approx -\Lambda\spd X/\ox\size=-X/\tcon$ from (unlinked) sweeps that start in between them.
In other words, we can approximate the motion by
 \begin{equation}
 dY_\tau = -\tcon^{-1}Y_\tau d\tau + 2\sqrt{\mig} dB_\tau,\label{brownian}
 \end{equation}
 where $B$ is a Brownian motion.
 We write $Y$ to emphasize that this is not quite the same as the real path of the lineages $X$.
In particular, unlike $X$, $Y$ does not include coalescence. 
(In two dimensions, $Y$ fails to approximate $X$ even when the lineages are just very close together,
but since most of the coalescence time will be spent at some distance away, it is still a useful approximation.)

We would like to find an explicit form for \eqref{ibdT}. 
To do this, we can rewrite in terms of the behavior of $Y$. 
First, note that the rate of coalescence for the two lineages when they are in the same place is $1/\pop$, and 
therefore the probability density of coalescence at time $\tau$ is $\approx  \frac{\delta(Y_\tau)}{\pop}\exp \left(-\int_0^\tau d\tau' \frac{\delta(Y_{\tau'})}{\pop}\right) $,
where $\delta$ is the Dirac delta.
(The exponential factor accounts for the possibility that the two lineages have already coalesced.)
Plugging this into \eqref{ibdT} gives:
\begin{align}
\ibd(x) = & \evo{e^{-2\mut \tcoal} \given X_0=x}{X}\nn\\
 \approx & \evo{\int_0^\infty d\tau \frac{\delta(Y_\tau)}{\pop}\exp\left(-2\mut\tau -\int_0^\tau d\tau' \frac{\delta(Y_{\tau'})}{\pop}\right) \given Y_0=x}{Y}.\label{ibdY}
\end{align}

We can use the Feynman-Kac formula (\cite{pham2009}, p25) to rewrite \eqref{ibdY} as an ordinary differential equation: 
\begin{equation}
0 =  2\mig \ibd'' + \left(2\mig\frac{d-1}{x}-\frac{x}{\tcon}\right)\ibd' - 2\mut \ibd +\frac{1}{\pop}\delta(x)(1-\ibd),\label{dibddx}
\end{equation}
where $\delta$ is the Dirac delta.
\eqref{dibddx} breaks down for $x\to0$ in $d=2$ dimensions; in this case, some kind of small-scale cutoff is needed,
 but this does not change the shape of  $\ibd(x)$ at larger scales \cite{barton2002a}.
 The last term in \eqref{dibddx} is just a boundary condition that sets the overall normalization of $\ibd$.
 
 The solution to \eqref{dibddx} can be written exactly in terms of special functions. (For $d=1$, \eqref{dibddx} is the Hermite equation, with solution
 $\ibd(x) \propto H_{-2\mut \tcon}\left(\frac{x}{2 \sqrt{\mig \tcon}}\right)$, where $H_\nu(z)$ is a Hermite function \cite{mathworldHermite}.)
However, approximate asymptotic solutions are more useful.
For $x \gg \xc \sqrt{2+4\mut\tcon}$, the dominant balance is $\ibd'\approx-2\mut\tcon\ibd/x$, with solution $\ibd \propto x^{-2\mut \tcon}$. 
For $x \ll \xc$, the pull of unlinked sweeps is negligible, and the solutions are close to the neutral solutions in \cite{barton2002a}.
At intermediate distances, there is a crossover regime where the form of the dependence of $\ibd$ on $x$ is independent of the mutation rate.

\subsection{Tightly-linked sweeps}

Above, we have focused on regions of the genome far from any recent sweeps. 
Ideally, however, we would like to be able to extend our analysis to include recently-swept regions. 
As a first approximation, we can say that the main effect of tightly-linked sweeps is that they can 
cause two widely-separated lineages to rapidly coalesce. 
The probability that a sweep recombining at rate $r$ with the focal neutral locus will
cause coalescence between two lineages separated by $x$ is $\approx \exp(-r x/\spd)/(1+2r\Upsilon)$,
where $\Upsilon$ is mean coalescence time for two lineages inside the wavefront of the sweep \cite{barton2013}.
We can therefore account for the effect of sweeps uniformly distributed over the genome 
by changing the coalescence kernel  in \eqref{ibdY} from $\delta(x)/\pop$ to 
\begin{align*}
p_\text{coal}(x) & \approx \delta(x)/\pop + \frac{2\Lambda}{\ox K}\int_0^\infty dr \frac{e^{-r x/\spd}}{1+2r\Upsilon}\\
& \approx \frac{2\Lambda}{\ox K}\frac{\spd}{x}\text{ for $x\gg\spd\Upsilon$}.
\end{align*}

For  $x\gg\spd\Upsilon$, \eqref{dibddx} then becomes 
\[
0 =  2\mig \ibd'' + \left(2\mig\frac{d-1}{x}-\frac{x}{\tcon}\right)\ibd' - 2\mut \ibd +\frac{2\Lambda}{\ox K}\frac{\spd}{x}(1-\ibd).
\]
For large $x$, there are two possible tail behaviors for the solution. 
If $2\mut\tcon<1$, then the pull of unlinked sweeps is strong enough that it is likely to bring lineages close together before 
they mutate, and $\ibd \propto x^{-2\mut \tcon}$ as above.
For $2\mut\tcon >1$, only recently-swept loci share recent enough ancestry to be likely to be identical in distant individuals,
and $\ibd \propto x^{-1}$.

\section{Simulation methods}

Forward-time simulations (purple histogram in \fig{ancdist}) were conducted using the algorithm from \cite{weissman2012}
(which draws on that of \cite{kim2003}),
modified so that population was subdivided into a line of $\size$ demes of $\pop$ individuals
each, with random dispersal between adjacent demes.
Because these simulations were extremely computationally demanding,
we also conducted approximate backwards-time simulations to get better statistics
and investigate rare events (blue histogram in \fig{ancdist}, and \figs{trace}{logancdist}).
These simulations followed a single lineage back in time at one neutral locus as it diffused through a continuous one-dimensional space.
Sweeps were treated as instantaneous events arising uniformly at random in space and time and across the genome.
Sweeps occurring at a recombination fraction $r$ from the focal locus pulled the lineage an exponentially-distributed distance with mean
$\spd/r$ or $\spd/(2r)$ (for $r < s$ and $r > s$, respectively), truncated at the origin of the sweep.
For both sets of simulations, the focal locus was at the center of a linear genome with map length $K$ Morgans.

\end{document}